# The future of census coverage surveys

### Kenneth Wachter[1]

*University of California, Berkeley*

**Abstract:** A quarter-century of statistical research has shown that census coverage surveys, valuable as they are in offering a report card on each decennial census, do not provide usable estimates of geographical differences in coverage. The determining reason is the large number of "doubly missing" people missing both from the census enumeration and from coverage survey estimates. Future coverage surveys should be designed to meet achievable goals, foregoing efforts at spatial specificity. One implication is a sample size no more than about 30,000, setting free resources for controlling processing errors and investing in coverage improvement. Possible integration of coverage measurement with the American Community Survey would have many benefits and should be given careful consideration.

## 1. Coverage surveys and their purposes

During the final decades of the Twentieth Century, proposals for the statistical adjustment of the decennial census in the United States provided a large-scale proving ground for approaches to the evaluation of statistical procedures. In many respects, David Freedman has been the central figure in these studies. He more than anyone else directed attention beyond political controversies surrounding particular decisions to the general scientific question of how one separates statistical findings driven by data from findings dependent on hypothetical assumptions in a modeling process.

Census adjustment is a topic well-suited to raise general questions about the nature and efficacy of models. First, the scale and complexity of the topic have meant that a wide range of statistical methods have played a role and come under scrutiny. Second, the professionalism of the Census Bureau and the willingness of Bureau leadership to make large data sets available to outside researchers have enabled independent replication of estimates to an extent rarely seen. Third, the background pressures of political struggle and legal review have kept the distinction between data and assumptions in the foreground.

A legacy of this enterprise of statistical examination is an unparalleled foundation for rational planning for census evaluation as we enter the Twentyfirst Century. In this essay, I bring together lessons that have been learned about the capabilities of census coverage surveys and their design and deployment for the future.

"Coverage" is a general demographic term referring to accuracy of enumeration in the face of errors of undercounting, overcounting, or miscounting members of a population. Census coverage surveys usually take the form of "post-enumeration" surveys conducted after the completion of a census enumeration. They have long accompanied census taking in statistically numerate societies.

[1]University of California, Department of Demography and Department of Statistics, 2232 Piedmont Avenue, Berkeley, California 94720-2120, USA, e-mail: wachter@demog.berkeley.edu

In the United States, census coverage surveys have been mounted in conjunction with every decennial census since 1950, with expanded efforts since 1980. The 1980 enterprise went under the acronym of PEP, for "Post-Enumeration Program". It made use, in part, of samples amounting to 168,000 households from the Current Population Survey rather than from its own field operation. The 1990 survey was called the PES, standing simply for "Post-Enumeration Survey". It included its own field survey with a sample, the "P-sample", of 5,290 block groups and about 170,000 housing units along with a corresponding sample of Census records, the "E-sample". The 2000 program was called ACE, for "Accuracy and Coverage Evaluation." As in 1990, it collected its own P-sample and E-sample, with an augmented sample size of about 300,000 housing units. Plans are in formation for a counterpart in 2010.

The original purpose of census coverage surveys is to generate indices of the likely extent of error in key census figures and of the uncertainties in population counts. In the United States, summary estimates of coverage error from each decennial survey have come to be regarded as a report card on the census. Receiving most attention have been the estimate of net national undercount and the estimates of differences between racial and ethnic groups. The difference in estimated net national undercount between African-Americans and members of other races is often referred to as "the differential undercount".

The special virtue of a full census enumeration is the refined information on geographical location which it provides. Surveys intrinsically have lower spatial resolution. A sample of a size adequate for precise estimates for large population aggregates cannot be indefinitely split up place by place to yield direct estimates for small geographical units. In the United States, census coverage surveys through the 1980 round were never intended to support disaggregated geographical estimates of undercounts.

From 1980 onward, however, census coverage surveys have been pressed into service for this originally unintended purpose, for estimates of spatial differences in coverage errors. Calculations were put in play for generating adjusted census counts area by area and ultimately block by block. Direct estimates being out of the question, a combination of data collection with extensive statistical modeling and inferences based on assumptions came to the fore.

In the to-and-fro of litigation, changing federal policies, and Supreme Court rulings, adjusted figures have been repeatedly computed, but they have never been accepted as official census counts. The 1980 PEP was redeployed to generate candidate sets of adjusted figures in response to a court decree. The 1990 PES was developed on a large scale for possible implementation of adjustment, and it was accompanied by a suite of evaluation projects known as the P-Studies and the E-Studies. Reports from these 1990 studies still constitute the most detailed information available for understanding coverage error and adjustment dynamics. The 2000 ACE, with its larger sample size, also included valuable evaluation studies, although high levels of initially undetected duplicate counting in the 2000 Census have complicated the investigations.

Many researchers have contributed to analyses of the surveys, estimates, and evaluation studies. Overviews with extensive references can be found in Freedman and Wachter [7], Anderson and Fienberg [1], and in Brown and Zhao [4] in this volume.

## 2. Lessons

Building on the exemplary technical literature, as it pertains to the subject of this essay, I offer a general conclusion:

> A quarter-century of experience tells us that census coverage surveys are useless for measuring spatial differences in coverage.

The purpose that can be met by census coverage surveys is their original purpose. They can serve to characterize overall levels of error and differentials between population groups and give a report card on the census. They cannot reliably ascertain geographical distributions.

Many difficulties and limitations of census coverage surveys with respect to geography have been identified and studied, but one is intractable and disposative. That one is the large "doubly missing" population.

The phrase "doubly missing" refers to people who are missing both from the Census enumeration and from estimates computed from the coverage survey. "Unreached people" is an alternate descriptor. The technical term for this kind of error is "correlation bias". The reason for correlation bias in survey-based estimates is obvious. People who are hard to count are also hard to survey.

The doubly missing add a "fifth cell" to the four cells of a two-way table used in the estimation process. Three cells contain observations: (1), at upper left, the number included in both census and survey, (2), at upper right, the number included in the census, missing from the survey, and (3), at lower left, the number missing from the census included in the survey. In the fourth, lower-right cell, it is customary to fill in the product $(4) = (2) \times (3)/(1)$. This product equals the number we should expect to see missing at random, in the counterfactual setting in which being missed in the survey were statistically independent of being missed in the census. Estimates from surveys routinely compensate for people missing at random but not for unmeasured correlations in missingness. The fifth cell represents the excess of people missing from both data systems due to correlation bias.

The estimation method in use with census adjustment is known as the "Capture-Recapture" or "Dual System Estimator", abbreviated DSE. It is described in Hogan [8] and Brown et al. [3] and in Brown and Zhao [4] in this volume. Dual System estimates are computed separately for broad population groups and then added together. The groups are called "post-strata" because individuals are assigned to them *post-hoc,* after data collection. Post-stratification controls for some dimensions of being hard to count. But reasons for being missed by census and survey takers are much more various and personal than are pinpointed by a classification into post-strata. Only a modest portion of correlation bias is removed by stratification exploiting available variables. Large excesses of unreached people remain missing beyond the allowances made by Dual System estimators.

We detect the existence of doubly missing people not from any information in the coverage survey, but from comparison with independent national population estimates by age, sex, and race from an approach known as Demographic Analysis. "Demographic Analysis" abbreviated DA, is written with capital letters D and A to distinguish it from generic analysis by demographers. DA draws on administrative records, birth and death registration, Medicare enrollments, and guesses at net immigration informed by ancillary survey-based estimates of the foreign-born [10]. DA is subject to many kinds of errors and uncertainties of its own, but it is the best available source for national numbers and it is the Census Bureau's standard against which other counts and estimates are compared.

Properties of the estimation methods and comparisons between Census, Dual System, and DA totals have been intensively scrutinized by the scientific community over the last twenty-five years. Evidence has accumulated about doubly missing people and their impact on estimates. Three conclusions bearing on the future of census coverage have emerged:

1. The doubly missing, for whom no geographical information is available, are so numerous that they amount to a substantial fraction of any census undercount.
2. Breakdowns of doubly missing people by sex and race show higher than average numbers among African-American males, in line with expectations.
3. There is strong circumstantial evidence that doubly missing people were unevenly distributed across regions of the country in 1990 and 2000.

## 3. Numbers of the doubly missing

We review these conclusions one by one, beginning with the numbers of doubly missing people. Three elements go into an estimate of doubly missing people, an initial DSE figure for net undercount from the coverage survey, an estimate of processing error in the DSE generated by the evaluation studies (P-studies and E-studies) and an estimate of total net undercount from Demographic Analysis. Overcounts as well as undercounts occur in Census figures. Net undercount is the amount by which undercounts exceed overcounts. It can be positive or negative. The magnitudes of estimates, choices among alternatives, and critiques of assumptions are set out in Wachter and Freedman [13] and Freedman and Wachter [7]. The summary here is based on these works, and readers should consult them for details.

The first and third of these elements, the DSE and DA, have already been mentioned. The second, processing error in the post-enumeration survey estimates, enters from a variety of sources. A prominent example is a failure in the 2000 survey to detect and correct for a large number of duplicate census records that independently came to light. Sources of processing error include geocoding errors, census day address errors, matching errors, imputation errors, undetected fabrications, imprecise matching of movers, and errors in balancing the scope of census and survey.

Many sources of processing error are measured by quality control procedures, blind recoding experiments, and targeted field followup of samples of cases. These measurements are used in estimates of overall correlation bias. The 1990 estimates of processing error were thoroughly vetted (Breiman [2], Fay and Thompson [6]). The 2000 estimates are subject to wider uncertainty (Freedman and Wachter [7]). Some components are positive, some negative. On balance, processing errors typically lead the DSE to overstate the net undercount and constitute a negative correction to the DSE figures. They are referred to as "measured biases" to distinguish them from correlation bias, which, in the context of quality control and field followup, is an unmeasured bias.

DSE figures for undercount come out at a percent or two of the whole population. Errors in survey processing come out at no less than a percent or two. Indeed, 98% or 99% accuracy is hard to achieve. Thus processing errors, concentrated among hard cases, can easily affect a sizable fraction of the undercount. It is not surprising that estimates of processing error are pivotal.

Magnitudes of the relevant quantities for 1990 drawn from Wachter and Freedman [13] and for 2000 drawn from Freedman and Wachter [7] are shown in Table 1.

Table 1
*Elements of coverage error, estimates in millions*

|                            | 1990 PES | 2000 ACE |
|----------------------------|----------|----------|
| Dual-System Estimate       | +5.3     | +3.3     |
| Processing Error           | −3.6     | −5.5     |
| Corrected Survey Estimate  | +1.7     | −2.2     |
| Doubly Missing People      | +3.0     | +2.5     |
| DA Estimate                | +4.7     | +0.3     |

The initial DSE figures are shown in the first row, estimates of processing error in the second row, and their difference, the corrected survey estimates, in the third row. The entries in the second row are middle choices from a range of alternatives presented in Wachter and Freedman [13].

For each census, the estimate of doubly missing people is the number, in the fourth row, that has to be added onto the corrected survey estimate to give the DA net undercount in the final row. The DA net undercount for 1990 is the difference between the DA population estimate of 253.394 million from the Office of the Secretary of Commerce [9] and the Census count of 248.710 million. For 2000, it is the corresponding difference $281.760 - 281.421$ million calculated from the revised DA estimate in Robinson et al. [10].

Table 1 indicates totals of doubly missing people in the millions, around 3 million in 1990, on the order of 2.5 million in 2000. In 1990 the doubly missing account for a majority of the net number of people missing from the Census according to the standard supplied by Demographic Analysis. In 2000, initial estimates suggested a small net Census overcount (attributed to duplicates) and revised estimates suggest a net undercount of a few hundred thousand. Doubly missing people offset the whole of the net undercount estimate derived from the coverage survey corrected for measured sources of processing error.

The numbers of doubly missing people are hefty in comparison to net undercounts. But they are not surprising when compared to the larger numbers of gross omissions. Gross omissions represent numbers of people omitted from the census before the numbers are offset by people erroneously enumerated in the census. Gross estimates are dependent on details of definition, but, at least by Census Bureau figures discussed in Freedman and Wachter [7], p. 10–11, the doubly missing in 2000 could be less than a quarter of overall gross omissions.

Net undercount comes, roughly speaking, by taking gross omissions and subtracting the large number of erroneous enumerations, including fabrications and duplications. People assigned to the wrong location of residence may be reckoned both as a gross omission omitted from their proper location and as an offsetting erroneous enumeration included at the incorrect location. While gross figures are interesting for judgments of face validity, the quantities relevant for geographical breakdowns are net undercounts.

The Dual System Estimator nets out the gross figures within each post-stratum, before the process called synthetic estimation which produces geographical estimates. Dual System estimates of coverage error for substantial areas are almost all positive. In 2000, all states and all but two congressional districts had positive estimates [7], p. 11. Offsetting overcounts and undercounts occur among post-strata and negatives along with positives are not rare for small jurisdictions and for blocks. But for units of the size for which coverage surveys give direct information, only positive numbers are at issue.

The DSE is distributing a stock of people (5.3 million in 1990) among geographical locations to be added to Census counts. Millions of these people are wrongly included in the stock, due to processing errors. Their locations are measured, but only with lower geographical resolution. Millions more, the doubly missing people, have to be added back in order to approximate true counts. Their locations are not measured at all.

## 4. Differentials and distributions

Along with information about numbers of doubly missing people, the accumulated evidence supports some generalizations about differentials and distributions.

Although Demographic Analysis only yields figures at the national level, it does break the national population down by age, sex, and race. The evaluation studies (henceforth abbreviated ES) for 1990 and 2000 have no breakdowns by age, but they do have breakdowns by sex and minority-non-minority status which can be converted to rough breakdowns by sex and race. With 1990 data, breakdowns from the PES, the ES, and DA have been combined to give approximations for numbers of doubly missing people by sex and two racial categories, African-Americans and Other Races. Awkward features of the 2000 post-stratification, which did not sharply separate males and females, have as yet discouraged parallel estimates for 2000.

Analysis in Wachter and Freedman [13], p. 199–200, indicates that the 1990 doubly missing people included about 5.6% of the African-American men in the population, about 1.0% of the African-American women, 1.2% of the men of other races, and about 0.6% of the women of other races. Men, about 2.2 million of them, outnumber women, about 0.8 million of them, by about 1.4 million. Demographic Analysis is thought to be especially accurate for African Americans because undocumented immigration contributes little to their numbers. The higher estimated rates of being doubly missing for African-American men are in line with expectations and help corroborate the calculations.

The Census Bureau, in conjunction with the evaluation studies in 1990 and 2000, presented estimates which were labeled "Correlation Bias" and were entered into the Bureau's Total Error Model. Examination of the calculations, however, shows that these Bureau figures were not estimates of total correlation bias, but rather of the excess of male over female correlation bias. A full discussion is given in Wachter and Freedman [13], p. 200–222. When this distinction is taken into account, the Bureau's estimates are in good agreement with the estimates cited here.

Unlike differentials for broad demographic subgroups, geographical distributions for doubly missing people are not amenable to calculation. That is the chief point about doubly missing people. They are missing from both census and survey, and information about their locations is not at hand.

There is, however, strong circumstantial evidence that the doubly missing people in 1990 and 2000 were very unevenly distributed across regions of the country. The evidence is set out in Wachter and Freedman [13], p. 202–207, Freedman and Wachter [7], p. 6–7, and Wachter [12], p. 110–113. In both 1990 and 2000 tabulations, states in the northeast and midwest with large central cities and heavy concentrations of African Americans have unexpectedly low DSE figures for net undercounts. The undercount differences between these states and those in the west are not fully explained by upstate areas with lower minority concentrations, and they show up sharply when metropolitan areas are compared. Since African-American men are over-represented among the doubly missing, it seems plausible that the doubly missing are over-represented in areas where African Americans are known to be numerous and where undercounts are unexpectedly low.

This circumstantial evidence applies at a very high level of geographical aggregation and it is at best suggestive. We cannot claim to know in any detail how the doubly missing are distributed, even by region, and certainly not by state or sub-state area. As shown in Wachter and Freedman [13], p. 205, the numbers of doubly missing are large enough to alter the qualitative pattern of DSE-based adjustments to the proportional shares of states in the national population. In brief, due to the doubly missing people, the coverage surveys are not providing meaningful information about geographical gradients in coverage.

Doubly missing people are only one of the problems that undermine geographical breakdowns. Doubly missing people affect geography at the level of broad regions and at the level of states. Efforts to carry down estimates to finer levels of geography, to congressional districts, cities, counties, and ultimately to blocks come up against the limitations of a "synthetic assumption" discussed in Brown et al. [3] and Wachter and Freedman [14]. At all geographical levels, coverage surveys are not suited for estimating geographical variations in coverage.

## 5. Samples for the future

Recognition that census coverage surveys cannot reliably measure variability from place to place has far-reaching implications for the design of future coverage surveys. The large sample sizes in 1990 and 2000 have been motivated by the ultimately unsuccessful attempts to generate geographically differentiated estimates. Restoring emphasis on the original purpose, the report-card function, foregoing geographical specificity, allows much smaller sample sizes to be sufficient.

The resources that would be set free by reductions in sample sizes involve savings of many tens of millions of dollars. A small part of the savings could be directed into coverage improvement, quality control, and field followup activities and bring major benefits to overall accuracy. In the past, random sampling errors, which are reduced by increases in sample size, have been dominated by systematic errors like the processing errors discussed in Section 3, which go under the heading of "non-sampling errors". Non-sampling errors are made harder to control and measure by increases in sample size. Smaller sample sizes yield dividends in quality as well as cost.

Specific recommendations about sample sizes need to be predicated on detailed calculations about optimal sample allocation and its interaction with choices about sampling strata as well as post-strata. A rough idea of appropriate sample sizes can, however, be gleaned from comparisons with the 2000 post-enumeration survey ACE. A motivation for the large sample size of about 300,000 housing units in ACE was an early aim to be in a position to provide direct survey-based estimates for state populations for congressional apportionment. Census Bureau calculations then determined that sampling errors could be kept within acceptable limits when the ACE sample was subdivided into as many as 448 post-strata.

The 2000 ACE post-stratification incorporates divisions by demographic groups and by geographical categories. The demographic breakdown uses 7 age-sex classes and 6 main categories of race and ethnicity. (The racial category of American Indians and Alaskan natives is further subdivided, by an extra criterion of location, between those on reservations and those off reservations.) Each of these demographic groups is subdivided to differing degrees among four regions, eight types of places, and among renters and owners, a distinction included for the sake of synthetic estimation for small areas. Unfortunately, one of the 7 age-sex classes merges males and females, making consistency checks and comparisons awkward. A balanced design would have had 8 age-sex classes. Details are shown in a table in Brown and Zhao [4] in this volume.

The demographic breakdowns for the 2000 ACE post-stratification, then, comprise essentially $8 \times 6 = 48$ main demographic groups. A sample size of $48/448$ or about a tenth of the ACE sample size would be more or less sufficient to meet the same targets for sampling error for demographic groups as were met in 2000. This line of reasoning suggests a sample size for future coverage surveys on the order of about $30,000$ housing units, that is to say, in the several tens of thousands, not in the hundreds of thousands.

A complementary calculation is given in Freedman and Wachter [7], p. 14–15, 22–23, 31. The calculation there takes into account within-poststratum heterogeneity across areas at levels inferred from studies of proxy variables. The calculation leads to suggested sample sizes a little less than $30,000$ housing units. Such a sample size would still allow separate estimates of coverage by type of place, by renter-owner status, by mail-back rate, and by other variables of interest. Marginal tabulations would remain usable. Only cross classifications, crossing such variables with demographic breakdowns, would be swamped by sampling error. Cross classifications serve little purpose in any case, once it is recognized that carrying down undercount estimates to small geographical areas is undermined by the scale of non-sampling errors and spatial heterogeneity, along with doubly missing people.

The most important tabulations from any coverage survey are those that can be combined with corresponding tabulations from demographic analysis. Age, sex, and race are paramount. A limitation of Demographic Analysis is the absence, as yet, of separate estimates for the Hispanic population and for Asian Americans. Some research has been underway at the Census Bureau regarding the feasibility of extending or supplementing DA with breakdowns for Hispanics based on other administrative records. The resources invested in DA by the Census Bureau have been suprisingly small, given the central role DA has played in 1990 and 2000. A couple of parts per hundred of the budgets invested in coverage surveys like ACE could readily and wisely be redirected into developing Demographic Analysis.

Reliable estimates of coverage from future coverage surveys depend not only on DA but also on estimates of processing error from successful evaluation studies, counterparts of the P-Studies and E-studies of 1990 and 2000. Smaller samples for the coverage survey itself should lead to tighter initial control of processing errors and should also allow larger budgets for field follow-up for evaluation. Evaluation studies of sufficient scope to support breakdowns of processing error by age as well as sex would facilitate tighter consistency checks on the characteristics of unreached people. Coverage surveys do not stand on their own. Their value comes from their integration with the broader data collection program, with the Census enumeration, the evaluation studies, and Demographic Analysis.

## 6. The American Community Survey

The suggestions made in Section 5 apply to any administrative setup for coverage surveys that the Census Bureau may adopt. More specific reflections arise with regard to the Bureau's major innovation for the beginning of the new century, the American Community Survey.

The inauguration of the American Community Survey (ACS) and the planned replacement of the census long form with ACS data offer a new alternative for census coverage measurement. Progress and plans for the ACS are described in U. S. Census Bureau [11]. At full implementation, the ACS will be surveying about 250,000 housing units per month. For the sake of continuity with the long form, ACS methods are aligned with traditional procedures for the decennial census. Mail-out and mail-back are supplemented with computer-assisted telephone interviews and personal interviews on a sample basis. Implementation has been postponed a few times, but it should arrive before 2010.

Residence rules in the ACS for assigning respondents to locations are based on a concept of "current residence" in contrast to the census definition of "usual residence". Principles and details are discussed in Cork and Voss [5]. At the time of the decennial census, it will be essential to have calibration studies which allow data users to translate between the residence concepts of the two sources. Consequently, there will be a pressing need for the temporary inclusion in the ACS of questions eliciting census-day residence. That need arises independent of any relation to census coverage measurement. However, these considerations open up an opportunity for replacing a separate census coverage survey in 2010 and beyond with a subsample of ACS returns from a period following the Census.

The 1980 PEP offers a precedent. It made use of two waves from the Current Population Survey (CPS) in place of a separate P-sample. The American Community Survey is much better suited than the CPS for such a role, because of the pre-existing alignment of its procedures with decennial census procedures. Benefits would also accrue directly to the ACS from a program for matching ACS responses to decennial Census records, given the importance of calibrating survey results against Census tabulations.

A limitation of the American Community Survey for the purpose of coverage measurement is the lower response rates that it achieves. The Census long form has always had lower response rates than the short form, and the ACS inevitably continues this pattern. Were non-response distributed at random, it would not affect results. The 1990 and 2000 post-enumeration surveys had response rates no better than the censuses. But, of course, non-response is not distributed at random. We expect it to be concentrated among the hard-to-count. Use of the ACS in place of a separate post-enumeration survey would not alleviate the problem of correlation bias, and non-response might well be larger than in a separate post-enumeration survey.

But with numbers of doubly missing people already in the millions, small increments or decrements in coverage hardly matter very much. The problem of correlation bias cannot be addressed in any case with incremental improvements in survey design. Credible estimates require combination of survey information with independent sources, primarily with Demographic Analysis, and call for a willingness to forego geographical breakdowns which the data cannot sustain.

Counterbalancing the limitations of the American Community Survey for coverage measurement, there are distinct advantages. As an ongoing enterprise, the American Community Survey is not subject to the massive logistical challenges of a one-off enterprise like the 1990 PES or the 2000 ACE. The survey staff is already trained and already experienced. Regional offices and systems for local operations are already in place. Bugs and kinks in procedures are already being resolved as they turn up. The prospects for reducing processing errors from sources of measured bias are therefore more favorable in the context of the American Community Survey than with a separate survey.

Each of the separate coverage surveys of the last several decades has brought surprises, forcing the Census Bureau into last-minute or after-the-fact regrouping. A computer-coding error in 1990, a court decision that derailed plans for sample-based Census followup for 2000, and an unanticipated number of undetected duplications in the 2000 enumeration are examples. The unexpected will never be wholly banished, but an established day-by-day functioning enterprise like the ACS is vulnerable to fewer logistic uncertainties than a large-scale separate operation mounted under pressures of time.

With richness of information comparable to the Census long form, the American Community Survey supplies a more secure basis for matching individuals to Census records than traditional coverage surveys. Matching error and errors of imputation of cases with unknown match status have been two serious sources of processing error identified by evaluation studies. With better initial matching, field follow-up can be concentrated where it is most needed. Furthermore, with the ACS, heavy investment of resources in field follow-up for the hard-to-find would be easier to justify, because of the double payoff for ACS data quality along with coverage assessment.

Accounting for movers has been a thorny part of coverage measurement. In-movers who arrive at an address in time for the survey but lived elsewhere on Census Day have to be matched to a Census record at a previous address to be confirmed as correct enumerations. Out-movers who are recorded in the Census but gone by survey date have to be distinguished from erroneous enumerations. The richer information about respondents in the ACS could facilitate mover matching.

Estimation for movers could benefit from the built-in continuity of American Community Survey operations over time. PES and ACE operations were themselves spread out over many months. Later data mainly came from harder cases. Duration since Census day was confounded with intrinsic difficulty of enumeration. The American Community Survey would furnish estimates for movers from ordinary, randomly-selected cases as a function of days since Census day. Such time series offer checks on estimates for movers. Presumably, the ACS will be suspended during the actual taking of the Census, so as to avoid direct interference with the enumeration. Some start-up effects associated with resumption would occur, but not on the scale associated with a separate coverage survey.

Integration of coverage measurement into the American Community Survey might have broader benefits for coverage improvement. The ACS will be continuously generating indicators of coverage difficulty for a sample of local areas over the stretch of time leading up to the Census enumeration. Measures like the rate of cases with "insufficient information" (II's) discussed in this volume by Brown and Zhao [4] have been used as proxies for net undercount in past studies of heterogeneity, Wachter and Freedman [14]. An ACS-based time series of such proxies could provide "leading indicators" of coverage risk and a basis for a fair and systematic allocation of special resources for intensive fieldwork.

Using data from the American Community Survey in place of a separate P-sample would only require a small ACS subsample, much less than a single month's stream of data. The ability to draw stratified subsamples from the ACS with over-sampling of key hard-to-count subgroups on the basis of already recorded characteristics would be a bonus.

In summary, whether or not coverage measurement is integrated into the American Community Survey in 2010 or in succeeding censuses, lessons from past experience should govern future planning. Census coverage surveys should be designed to do well what they can do well. Smaller samples and better integration carry many benefits. Savings in costs can be directed into intensive matching and into evaluation follow-ups. Under any foreseeable arrangement, doubly missing people will still be preventing reliable geographical breakdowns for undercount. But better control over measured biases can allow more precise comparisons with results from Demographic Analysis. The result can be superior estimates of coverage for population subgroups, classified by age, sex, and race, and a more cogent report card for the evolving decennial census.